\documentclass[10pt,twoside]{article}
\usepackage{Latex-document}
\usepackage{multicol}
\usepackage{graphicx}
\newtheorem{theorem}{Theorem}
\newtheorem{definition}[theorem]{Definition}

\def\volumeno{I}
\title{\bf On the Work of Madhu Sudan: \vskip -2mm the 2002 Nevalinna Prize Winner\vskip 6mm}
\author{Shafi Goldwasser\thanks{Weizmann Institute of Science, Israel
and Massachusetts Institute of Technology, USA}\vspace*{-0.5cm}}
\date{\vspace{-8mm}}

\begin{document}

\maketitle \thispagestyle{first} \setcounter{page}{105}

\vskip 12mm

\section{Introduction}

\vskip-5mm \hspace{5mm}

Madhu Sudan's work spans many areas of computer science theory including
computational complexity theory, the design of efficient algorithms,
algorithmic coding theory, and the theory of program checking and
correcting.

Two results of Sudan stand out in the impact they have had on the
mathematics of computation. The first work shows a probabilistic
characterization of the class NP -- those sets for which short and
easily checkable proofs of membership exist, and demonstrates
consequences of this characterization to classifying the complexity of
approximation problems. The second work shows a polynomial time
algorithm for list decoding the Reed Solomon error correcting codes.

This short note will be devoted to describing Sudan's work on
probabilistically checkable proofs -- the so called {\it PCP theorem}
and its implications. We refer the reader to \cite{Sudan:complexity,
sudan-sigact} for excellent expositions on Sudan's breakthrough work on
list decoding,  and its impact on the study of computational aspects of
coding theory as well as the use of coding theory within complexity
theory.

Complexity theory is concerned with how many resources such as time and
space are required to perform various computational tasks. Computational
tasks arise in classical mathematics as well as in the world of computer
science and engineering. Examples of what we may call a computational
task include finding a proof for a mathematical theorem, automatic
verification of the correctness of a given mathematical proof, and
designing algorithms for transmitting information reliably through a
noisy channel of communication. Defining what is a `success' when
solving some of these computational tasks is still a lively and
important part of research in this stage of development of complexity
theory.

A large body of Sudan's work, started while he was working on his PhD
thesis, addresses the automatic verification of the correctness of
mathematical proofs. Many issues come up: how should we encode a
mathematical proof so that a computer can verify it, which mathematical
statements have proofs which can be quickly verified, and what is the
relation between the size of the description of the theorem and the size
of its shortest proof which can be quickly verified. The work of Sudan
sheds light on all of these questions.

\section{Efficient proof checking}

\vskip-5mm \hspace{5mm}

Let us start with the classic notion of efficiently checkable proofs,
which goes back to the early days of computer science in the early
seventies when the NP class was defined \cite{Cook,Levinnp}.

\begin{definition}
The class $NP$ consists of those sets $L\subset \{0,1\}^*$ for which
there exists polynomial time verification algorithm $V_L$ and polynomial
$p$ such that $x\in L$ if and only if there exists a
$y_x\in{\{0,1\}^{p(|x|)}}$ which makes $V_L(x,y_x)=TRUE$. We call $V_L$
the {\it NP-verifier} for the language $L\in NP$, and $y_x$ the {\it
NP-witness} for $x$ in $L$.
\end{definition}

One example of $L\in NP$ is the set of pairs $(G,k)$ where $k\in {\bf
Z}$ and $G$ is a graph which contain a complete subgraph on $k$ vertices
-- the so called CLIQUE problem. The NP-witness for pair $(G,k)\in
CLIQUE$ is the complete subgraph in $G$ of size $k$. Another example is
the set of all logical formulas for which a truth assignment to its
Boolean variables exists which makes it true -- the SATISFIABILITY
problem. The NP-witness for a logical formula $\phi$ is a particular
setting of its variables which make the formula satisfiable. Graphs,
logical formulas, and truth assignments can all be encoded as binary
strings.

\section{Probabilistic checking of proofs}

\vskip-5mm \hspace{5mm}

In the eighties, extensions of the notion of an efficiently verifiable
proof were proposed to address issues arising in disciplines involving
interactive computation such as cryptography. The extensions incorporate
the idea of using randomness in the verification process and allow a
negligible probability of error to be present in the verification
process. Variants of probabilistic proof systems include interactive
proofs \cite{GMR}, public-coin interactive proofs \cite{babai},
computational arguments\cite{BC}, CS-proofs \cite{Micalics}, Holographic
proofs \cite{BLFS}, multi-prover interactive proofs \cite{BGKW},
memoryless oracles \cite{FRS}, and probabilistically checkable proofs
\cite{FGLSS,AS}. The latter three definitions are equivalent to each
other although each was introduced under a different name.

By the early nineties probabilistically checkable proofs proofs were
generally accepted as the right extension for complexity theoretic
investigations. The class PCP of sets for which membership can be
checked by "probabilistically checkable proofs" is defined as follows.

\begin{definition}
Let $L\subset\{0,1\}^*$. For $L$ in PCP, there exists a probabilistic
polynomial time verification algorithm $V_L$
\begin{itemize}
\item
if $x\in L$, then there exists a $O_x\in\{0,1\}^*$ such that
$Prob[V_L^{O_x}(x) = TRUE] > 1$
\item
if $x\notin L$, then for all $O_x\in \{0,1\}^*$, $Prob[V_L^{O_x}(x) =
TRUE] <{1\over 2}  $.
\end{itemize}
The probabilities above are taken over the probabilistic choices of the
verification algorithm $V_L$. The notation $V_L^{O_x}$ means that $V_L$
does not receive $O_x$ as an input but rather can read individual bits
in $O_x$ by specifying their locations explicitly. We call $V_L$ the
PCP-verifier for $L\in PCP$, and $O_x$ the PCP-witness for $x$ in $L$.
\end{definition}

A few comments are in order.

For each bit of $O_x$ read, we charge $V_L$ for the time it takes to
write down the address of the bit to be read. The requirement that $V_L$
runs in polynomial time implies then that the length of the PCP-witness
for $x$ is bounded by an exponential in $|x|$.

A verifier may make an error and accept incorrectly, but the probability
of this event can be made exponentially (in $|x|$) small by running a
polynomial number of independent executions of $V_L$ and accepting only
if all executions accept. In light of the above, we argue that
probabilistically checkable proofs capture what we want from any
efficiently checkable proof system: correct statements are always
accepted, incorrect statement are (almost) never accepted, and the
verification procedure terminates quickly.

Are probabilistically checkable proofs more powerful than the
deterministic NP style proofs? Developments made in a sequence of
beautiful papers \cite{shamir, LFKN, BFL}, finally culminated in the
result of Babai et. al. \cite{BFL} showing that indeed
$PCP=NEXPTIME$.\footnote{ The class $NEXPTIME$ is defined exactly in the
same manner as $NP$ except that the verifier $V_L$ has exponential time
and the witness may be exponentially long. } By the separation of the
non-deterministic time hierarchy, it is known that $NP$ is strictly
contained in $NEXPTIME$. Thus indeed, the probabilistic checking of
proofs is more powerful than the classical deterministic one (at least
when the verifier is restricted to polynomial time).

Soon after the power of PCP verifiers was characterized, a finer look
was taken at the resources PCP verifiers use. Two important resources in
classifying the complexity of language $L$ were singled out
\cite{FGLSS}:  the amount of randomness used by the PCP verifier and the
number of bits it reads from the PCP-witness (the latter number is
referred to as the {\it query size} of $V_L$).

\begin{definition}
Let $PCP(r(n),q(n))$ denote class of sets $L\in PCP$ for which there
exists a PCP verifier for $L$ which on input $x\in \{0,1\}^n$ uses at
most $O(r(n))$ random bits and reads at most $O(q(n))$ bits of the
witness oracle $O_x$. \footnote{$O(g(n)=cf(n)$ s.t. there exists a
constant $c$ such that $g(n)\leq cf(n)$ for all $n$ sufficiently
large$\}$}
\end{definition}

Obviously, $NP\subset PCP(0,\cup_c{n^c})$ as an NP verifier is simply a
special case of the PCP verifier which does not use any randomness.
Starting with scaling down the result of \cite{BFL} it was shown (or at
least implied ) in a sequence of improvements \cite{BLFS,FGLSS,AS} that
$NP\subset PCP(\log n, poly (\log n))$. These results successively
lowered the number of bits that the PCP- verifier needs to read from the
PCP-witness, but it seemed essential for the correctness of the
verification procedure that this number should be a function which grows
with the size of the
input.

In the eighties, extensions of the notion of an efficiently verifiable
proof were proposed to address issues arising in disciplines involving
interactive computation such as cryptography. The extensions incorporate
the idea of using randomness in the verification process and allow a
negligible probability of error to be present in the verification
process. Variants of probabilistic proof systems include interactive
proofs \cite{GMR}, public-coin interactive proofs \cite{babai},
computational arguments\cite{BC}, CS-proofs \cite{Micalics}, Holographic
proofs \cite{BLFS}, multi-prover interactive proofs \cite{BGKW},
memoryless oracles \cite{FRS}, and probabilistically checkable proofs
\cite{FGLSS,AS}. The latter three definitions are equivalent to each
other although each was introduced under a different name.

By the early nineties probabilistically checkable proofs proofs) were
generally accepted as the right extension for complexity theoretic
investigations. The class PCP of sets for which membership can be
checked by "probabilistically checkable proofs" is defined as follows.

\begin{definition}
Let $L\subset\{0,1\}^*$. For $L$ in PCP, there exists a probabilistic
polynomial time verification algorithm $V_L$
\begin{itemize}
\item
if $x\in L$, then there exists a $O_x\in\{0,1\}^*$ such that
$Prob[V_L^{O_x}(x) = TRUE] > 1$
\item
if $x\notin L$, then for all $O_x\in \{0,1\}^*$, $Prob[V_L^{O_x}(x) =
TRUE] <{1\over 2}  $.
\end{itemize}
The probabilities above are taken over the probabilistic choices of the
verification algorithm $V_L$. The notation $V_L^{O_x}$ means that $V_L$
does not receive $O_x$ as an input but rather can read individual bits
in $O_x$ by specifying their locations explicitly. We call $V_L$ the
PCP-verifier for $L\in PCP$, and $O_x$ the PCP-witness for $x$ in $L$.
\end{definition}

A few comments are in order.

For each bit of $O_x$ read, we charge $V_L$ for the time it takes to
write down the address of the bit to be read. The requirement that $V_L$
runs in polynomial time implies then that the length of the PCP-witness
for $x$ is bounded by an exponential in $|x|$.

A verifier may make an error and accept incorrectly, but the probability
of this event can be made exponentially (in $|x|$) small by running a
polynomial number of independent executions of $V_L$ and accepting only
if all executions accept. In light of the above, we argue that
probabilistically checkable proofs capture what we want from any
efficiently checkable proof system: correct statements are always
accepted, incorrect statement are (almost) never accepted, and the
verification procedure terminates quickly.

Are probabilistically checkable proofs more powerful than the
deterministic NP style proofs? Developments made in a sequence of
beautiful papers \cite{shamir, LFKN, BFL}, finally culminated in the
result of Babai et. al. \cite{BFL} showing that indeed
$PCP=NEXPTIME$.\footnote{ The class $NEXPTIME$ is defined exactly in the
same manner as $NP$ except that the verifier $V_L$ has exponential time
and the witness may be exponentially long. } By the separation of the
non-deterministic time hierarchy, it is known that $NP$ is strictly
contained in $NEXPTIME$. Thus indeed, the probabilistic checking of
proofs is more powerful than the classical deterministic one (at least
when the verifier is restricted to polynomial time).

Soon after the power of PCP verifiers was characterized, a finer look
was taken at the resources PCP verifiers use. Two important resources in
classifying the complexity of language $L$ were singled out
\cite{FGLSS}:  the amount of randomness used by the PCP verifier and the
number of bits it reads from the PCP-witness (the latter number is
referred to as the {\it query size} of $V_L$).

\begin{definition}
Let $PCP(r(n),q(n))$ denote class of sets $L\in PCP$ for which there
exists a PCP verifier for $L$ which on input $x\in \{0,1\}^n$ uses at
most $O(r(n))$ random bits and reads at most $O(q(n))$ bits of the
witness oracle $O_x$. \footnote{$O(g(n)=cf(n)$ s.t. there exists a
constant $c$ such that $g(n)\leq cf(n)$ for all $n$ sufficiently
large$\}$}
\end{definition}

Obviously, $NP\subset PCP(0,\cup_c{n^c})$ as an NP verifier is
simply a special case of the PCP verifier which does not use any
randomness. Starting with scaling down the result of \cite{BFL} it
was shown (or at least implied ) in a sequence of improvements
\cite{BLFS,FGLSS,AS} that $NP\subset PCP(\log n, poly (\log n))$.
These results successively lowered the number of bits that the PCP
verifier needs to read from the PCP-witness, but it seemed
essential for the correctness of the verification procedure that
this number should be a function which grows with the size of the
input.

\section{The PCP theorem}

\vskip-5mm \hspace{5mm}

In a breakthrough,  which has since  become known as {\it the PCP
theorem}, Sudan and his co-authors characterized the class $NP$ exactly
in terms of $PCP$. They showed that $NP$ contains exactly those
languages in which a $PCP$-verifier can verify membership using {\it
only} a constant query size and using logarithmic (in the instance size)
number of coins. More over, there exists a polynomial time procedure to
transform an $NP$-witness of $x$ in $L$ into a PCP-witness of $x$ in
$L$.

\begin{theorem}\cite{ALMSS}
$NP =PCP (\log n, 1)$
\end{theorem}

On an intuitive level,  the PCP theorem says that there exist a
probabilistic verifier for proofs of mathematical assertions which can
look only at a constant number of bit positions at the proof and yet
with some positive probability catch any mistake made in a fallacious
argument.

The proof of the PCP theorem is deep, beautiful, and quite complex. It
brings together ideas from algebra, error correcting codes,
probabilistic computation, and program testing.

Although the PCP theorem establishes a complexity result, its proof is
algorithmic in nature, as it is a transformation of an NP-witness and a
deterministic NP-verifier for $L\in NP$ into a PCP-witness and an
PCP-verifier for $L$. As such it uses methods from the design of
computer algorithms and the design of error correcting codes. Several
excellent expositions of the proof appeared \cite{Sudan:pcp}.

In a very strong sense, the act of transforming an NP witness into a PCP
witness is similar to transforming a message into an error correcting
code word. The analogy being that a code word is an encoding of a
message which enables error detection in spite of noise injected by an
adversary, and a PCP witness is an encoding of a proof which enables
detection with high probability of an error in spite the best efforts to
hide it made by a cheating pretend-to-be prover.

Yet, the act of classic {\it decoding} of a code word is very different
than the act of checking the correctness of a PCP witness . Whereas in
error correcting codes one attempts to recover the entire original
message from the corrupted  code word if too much noise has not
occurred; here we only want to verify that the PCP-witness is a proper
encoding of a valid NP-witness (of the same fact) which would have
convinced an NP-verifier to accept. It suffices to read only a constant
number of bit positions to achieve the latter task, whereas the decoding
task depends on reading the entire code word.

One of the subsequent contributions of Sudan, involves constructing a
new type of {\it locally testable codes} \cite{SudanFriedl,
SudanGoldreich}. Locally testable codes are error-correcting codes for
which error detection can be made with probability proportional to the
distance of the non-codeword from the code, based on reading only a
constant number of (random) symbols from the received word. A related
concept is that of {\it locally decodable codes}
\cite{Trevisan1,Trevisan2} which are error correcting codes which may
enable recovery of part of the message (rather than the entire message)
by reading only part of the corrupted code word.

\section{PCP and hardness of approximation}

\vskip-5mm \hspace{5mm}

The intellectual appeal of the PCP theorem statement is obvious. What is
much less obvious and what has been the main impact of the PCP theorem
is its usefulness in proving NP hardness of many approximate versions of
combinatorial optimization problems. A task which alluded the
theoretical computer science community for over twenty years.

Shortly after the class $NP$ and the companion notion of an
$NP$-complete and NP-hard problems\footnote{ A set is ${\cal NP}$-hard
if {\em any} efficient algorithm for it, can be used to efficiently
decide every other set in ${\cal NP}$. An NP set which is NP-hard is
called NP-complete.  By definition, every ${\cal NP}$-complete language
is as hard to compute  as any other. } were introduced, Karp illustrated
its great relevance to combinatorial optimization problems in his 1974
paper \cite{Karp72}, He showed that a wide collection of optimization
problems ( including the minimum travelling salesman problem in a graph,
integer programming, minimum graph coloring and maximum graph clique
suitably reformulated as language membership problems) are NP-complete.
Proving that a problem is NP-complete is generally taken to mean that
they are intrinsically intractable as otherwise $NP=P$.

In practice this means there is no point in wasting time trying to
devise efficient algorithms for NP-complete problems, as none exists
(again if $NP\neq P$). Still these  problems do come up in applications
all the time, and need to be addressed. The question is,  how? Several
methods for dealing with NP-completeness arose in the last 20 years.

One technique is to devise algorithms which provably work efficiently
for particular input distribution on the instances (``average''
instances ) of the NP-complete problems. It is not clear however how to
determine whether your application produces such input distribution.

Another direction has been to devise approximation algorithms. We say
that an approximation algorithm $\alpha$-approximates a maximization
problem if, for every instance, it provably guarantees a solution of
value which is at least $1\over{\alpha}$ of the value of an optimal
solution; an approximation algorithm is said to $\alpha$-approximate a
minimization problem if it guarantees a solution of value at most
$\alpha$ of the value of an optimal solution.

Devising approximation algorithms has been an active research area for
twenty years, still for many NP-hard problems success has been illusive
whereas for others good approximation factors were achievable. There has
been no theoretical explanation of this state of affairs. Attempting to
prove that approximating the solution to NP-hard problems is in itself
NP-hard were not successful.

The PCP theorems of Sudan and others, starting with the work of Feige
et. al. \cite{FGLSS}, has completely revolutionized this state of
affairs. It is now possible using the PCP characterization of NP to
prove that approximating many optimization problems each for different
approximation factors is in itself NP-hard. The mysteries of why it is
not only hard to solve optimization problems exactly but also
approximately, and why different NP-hard problems behave differently
with respect to approximation have been resolved.

The connection between bounding the randomness and query complexity of
PCP-verifiers for NP languages and proving the NP hardness of
approximation was established in \cite{FGLSS,AS} for the Max-CLIQUE
problem (defined below). It seemed at first like an isolated example.
The great impact of Sudan et. al.'s \cite{ALMSS} theorem was in showing
this was not the case. They showed that proving characterization of NP
as $PCP(log n, 1)$ implies the NP hardness of approximation for a
collection of NP-complete problems including Max-3-SAT, Max-VERTEX
COVER, and others (as well as improving the Max-CLIQUE hardness factor).

The basic idea is the following: A  PCP type theorem provides a
natural(?) optimization problem which cannot be efficiently approximated
by any factor better than 2 as follows. Fix a PCP-verifier $V_L$ for an
$NP$ language $L$ and an $x$. Any candidate PCP-witness $O_x$ for $x$
defines an acceptance probability of $V_L^{O_x}(x)$. The gap of $1/2$ in
the maximum acceptance probability for $x\in L$ versus $x\notin L$
(which exists by the definition of $PCP$) implies that it is NP-hard to
2-approximate the maximum acceptance probability of $V_L$. In other
words, the existence of a  polynomial time algorithm to 2-approximate
the acceptance probability of $x$ by $V_L$ would imply that $NP=P$.

For different optimization problems, showing hardness of approximation
is done by demonstrating {\it reductions } from variants of the above
optimization problem. These reductions are far more complex than
reductions showing NP-hardness for exact problems as one needs to
address the difference in in-approximability  factors of problems being
reduced to each other.

Moreover, these new NP-hardness results have brought on a surge of new
research in the algorithmic community as well. New approximation
algorithms have been designed which at times have risen to the task of
meeting from above the approximation factors which were proved using PCP
theorems to be best possible (unless $NP=P$) . This has brought on a
meeting of two communities of researchers: the algorithm designers and
complexity theorists. The former may take the failure of the latter to
prove NP hardness of approximating a problem within a particular
approximation factor as indication of what factor is feasible and vice
versa.

This radical advance is best illustrated by way of a few examples.
Finding the exact optimal solution to all of the following problems is
${\cal NP}$-complete. Naive approximation algorithms existed for a long
time, which no one could improve. They yield completely different
approximation factors. For some of these problems we now have
essentially found optimal approximation problem. Any further advancement
will imply that $NP$ problems are efficiently solvable.

\noindent{\bf Max-CLIQUE}: Given a finite graph on $n$ vertices, find
the size of the largest complete subgraph. A single vertex solution is
within factor $n$ of optimal. More elaborate algorithms give factor
$n^{.999}$. This problem was the first one to be proved hard to
approximate using PCP type theorem \cite{FGLSS}. It is now known that
achieving a factor of $n^{1-\epsilon}$ is ${\cal NP}$-hard for every
$\epsilon
>0$~\cite{Hastad:clique}.

\noindent{\bf Max-3-SAT}: Given a logical formula in conjunctive normal
form with $n$ variables where there is at most 3 literals per clause,
determine the maximal number of clauses which  can be satisfied
simultaneously by a single truth assignment. A simple probabilistic
algorithm satisfies $1\over 2$ of the clauses. It is now known
\cite{Hastad:opt} that achieving a factor $7/8 - \epsilon$ for $\epsilon
> 0$ approximation factor is $NP$-hard even if the formula is
satisfiable. At the same time \cite{zwick} has shown an algorithm which
matches the $7/8$ approximation factor when the formula is satisfiable.

\noindent {\bf Min-Set Cover}: Given a collection of subsets of a given
finite universe of size $n$, determine the size of the smallest
subcollection that covers every element in the universe. A simply greedy
algorithm, choosing the subsets which maximizes the coverage of as many
yet uncovered elements as possible, yields a factor $\ln n$ from
optimal. It is now known that approximation by a factor of $(1-\epsilon)
\ln n$ is ${\cal NP}$-hard for every $\epsilon
>0$~\cite{Feige:Setcover}.

We point the reader to a collection of papers and expositions by Sudan
himself \cite{sudan-webpage} on these works as well as exciting further
developments.

\bibliographystyle{plain}

\newpage

\title{\vspace*{12mm} \centerline{\Large \bf Madhu Sudan}\vskip 6mm}
\author{\normalsize Associate Professor of Electrical Engineering and Computer Science, MIT}
\date{}

\maketitle

\vskip 12mm

\begin{tabular}{p{2.5cm}p{8cm}}
  1987 & B.Tech. in Computer Science, Indian Institute of Technology at New
  Delhi \\
  1992 & Ph.D. in Computer Science, University of California at Berkeley
  \\
  1992--1997 & Research Staff Member, IBM Thomas J. Watson \\
  & Research Center Mathematical Sciences Department \\
  1997 & Associate Professor, Department of Electrical Engineering and Computer
  Science at Massachussetts Institute of Technology
\end{tabular}

\vskip 5mm

\noindent {\bf Areas of Special Interests}

Theoretical Computer Science, Algorithms, Computational Complexity,
Optimization, Coding Theory.

\vfill

\begin{center}
\includegraphics[scale=0.7]{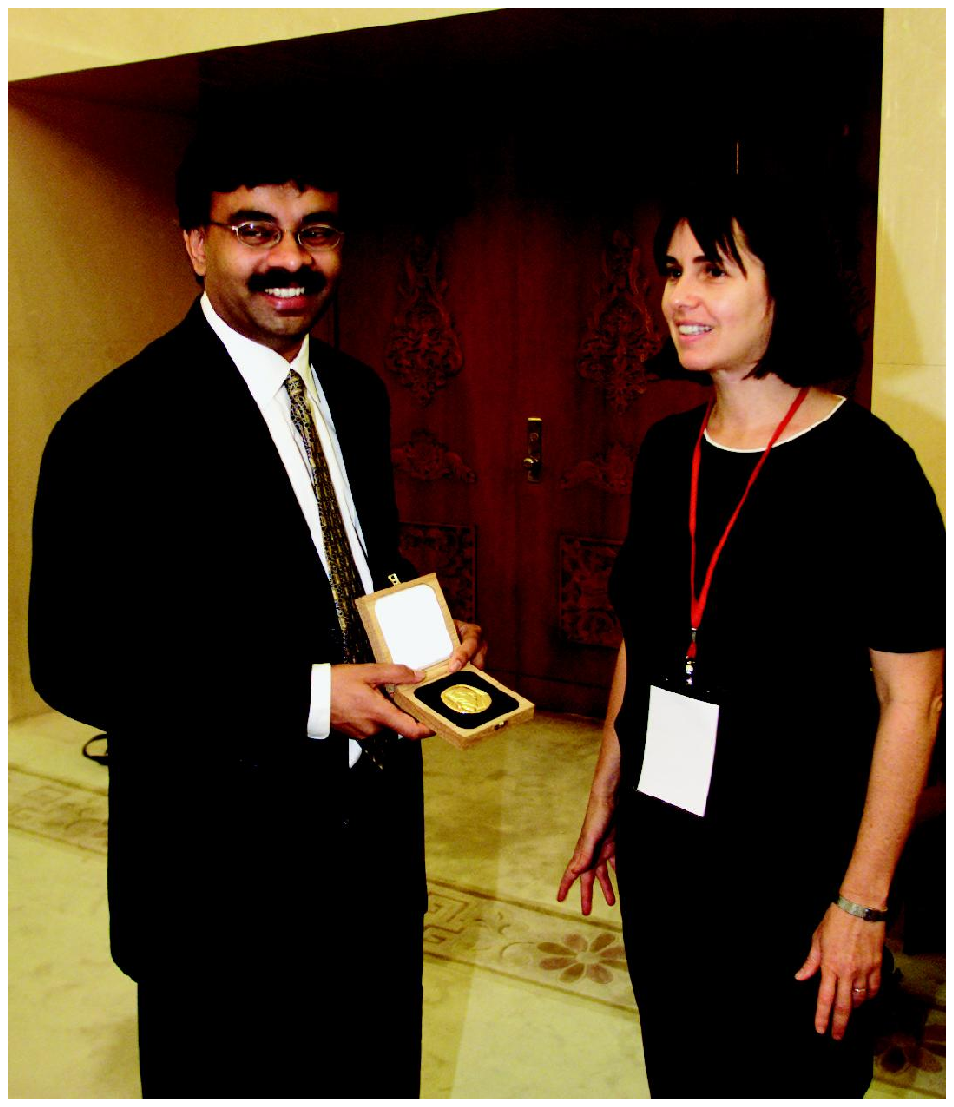}

\bigskip

M. Sudan (left) and S. Goldwasser
\end{center}

\label{lastpage}

\end{document}